\documentclass[sn-nature]{sn-jnl_edited}% referee option is meant for double line spacing

\setcitestyle{super}
\setcitestyle{super,open={},close={}}
\usepackage{setspace}

\usepackage{graphicx}%
\usepackage{multirow}%
\usepackage{amsmath,amssymb,amsfonts}%
\usepackage{amsthm}%
\usepackage{mathrsfs}%
\usepackage[title]{appendix}%
\usepackage{xcolor}%
\usepackage{textcomp}%
\usepackage{manyfoot}%
\usepackage{booktabs}%
\usepackage{algorithm}%
\usepackage{algorithmicx}%
\usepackage{algpseudocode}%
\usepackage{listings}%

\usepackage{siunitx}
\usepackage{braket}
\usepackage{upgreek} 
\usepackage{printlen}
\usepackage[normalem]{ulem}

\usepackage[dvipsnames]{xcolor}

\newcommand{\bmheadNoDot}[1]{%
  \par\smallskip
  \noindent\textbf{#1}\hspace{0.1em}%
}
\makeatletter
\def\ps@plain{%
  \let\@mkboth\@gobbletwo%
  \let\@oddhead\@empty\let\@evenhead\@empty%
  \def\@oddfoot{\vbox to 18pt{\vfill\reset@font\rmfamily\hfil\thepage\hfil}}% Entfernt 'ddd'
  \let\@evenfoot\@oddfoot%
}
\makeatother
\theoremstyle{thmstyleone}%
\theoremstyle{thmstyletwo}%

\theoremstyle{thmstylethree}%

\begin{document}

\title[Article Title]{Laser-enhanced quantum sensing boosts sensitivity and dynamic range}

\author*[1]{\fnm{Floian} \sur{Schall}}\email{florian.schall@iaf.fraunhofer.de}
\author[1]{\fnm{Lukas} \sur{Lindner}}\nomail
\author[1,2]{\fnm{Yves} \sur{Rottstaedt}}\nomail
\author[1]{\fnm{Marcel} \sur{Rattunde}}\nomail
\author[1]{\fnm{Florentin} \sur{Reiter}}\nomail
\author[1]{\fnm{R\"udiger} \sur{Quay}}\nomail
\author[3]{\fnm{Roman} \sur{Bek}}\nomail
\author[4]{\fnm{Alexander M.} \sur{Zaitsev}}\nomail
\author[5,6]{\fnm{Takeshi} \sur{Ohshima}}\nomail
\author[7]{\fnm{Andrew D.} \sur{Greentree}}\nomail
\author*[1]{\fnm{Jan} \sur{Jeske}}\email{jan.jeske@iaf.fraunhofer.de}

\affil[1]{\orgname{Fraunhofer Institute for Applied Solid State Physics IAF}, \orgaddress{\street{Tullastrasse 72}, \city{Freiburg}, \postcode{79108}, \country{Germany}}}

\affil[2]{\orgdiv{Institute for Theoretical Physics}, \orgname{University Leipzig}, \orgaddress{\street{Br\"uderstrasse 16}, \city{Leipzig}, \postcode{04103}, \country{Germany}}}

\affil[3]{\orgname{Twenty-One Semiconductors GmbH}, \orgaddress{\street{Kiefernweg 4}, \city{Neckartenzlingen}, \postcode{72654}, \country{Germany}}}

\affil[4]{\orgname{College of Staten Island (CUNY)}, \orgaddress{\street{Victory Blvd}, \city{Staten Island}, \postcode{10312}, \state{New York}, \country{USA}}}

\affil[5]{\orgname{National Institutes for Quantum Science and Technology (QST)}, \orgaddress{\street{1233 Watanuki}, \city{Takasaki}, \postcode{370-1292}, \state{Gunma} \country{Japan}}}

\affil[6]{\orgdiv{Department of Materials Science}, \orgname{Tohoku University}, \orgaddress{\street{Aoba}, \city{Sendai}, \postcode{980-8579}, \state{Miyagi} \country{Japan}}}

\affil[7]{\orgdiv{ARC Centre of Excellence for Nanoscale BioPhotonics, School of Science}, \orgname{RMIT University}, \orgaddress{\city{Melbourne}, \postcode{VIC 3001}, \country{Australia}}}

\abstract{Magnetometers based on nitrogen-vacancy (NV) centers in diamond have emerged as the most important solid-state quantum sensors. However, ensembles are limited in optical contrast to typically a few percent and high-sensitivity variants usually possess only a few $\mathrm{\upmu}$T dynamic range. Here, we demonstrate a laser threshold magnetometry-based NV system that avoids these limitations. By integrating the NV centers into a laser cavity and showing magnetic-field-dependent shifts of the laser threshold, we observe 100\,\% optical contrast, i.\,e., we are able to entirely switch off the laser system with the NV centers magnetic resonance. At the same time we achieve strong output signals up to 50\,mW. The system exhibits a photon-shot-noise-limited (PSNL) sensitivity of $<$400\,fT/$\sqrt{\textrm{Hz}}$ for all vector components, which we demonstrate to improve super-linearly with contrast. The ratio of the sensing-relevant parameters PSNL sensitivity and dynamic range, that can be traded at the cost of each other, marks an improvement factor of up to 590 over typical fluorescence-based readout and vapor cell sensors while also adding vector magnetometry capabilities. Such performance improvements provide a perspective for a highly sensitive magnetometer, which could be operated outside a magnetically-shielded room. This could bring a new generation of sensors for applications
including magnetoencephalography, magnetic navigation, and magnetic anomaly detection.}




\maketitle
%Gesamt Textbreite: \printlength{\textwidth} \\
%Spaltenbreite: \printlength{\columnwidth}
%\doublespacing
Accurate measurements of magnetic fields are essential in diverse areas, from biomedical diagnostics \cite{boto_meg, brookes_meg, aslam_meg, morales_mcg} to navigation \cite{searcy_navigation, afzal_navigation, wang_navigation} and resource exploration \cite{schmidt_mining, collar_mining}. Furthermore in health care, magnetic field sensing bears great potential for improvements over established electric measurements \cite{boto_meg, schofield_meg_eeg, fu_review, budker_review}. However, while state-of-the-art magnetometers such as superconducting quantum interference devices (SQUIDs) and vapor-cell sensors offer excellent sensitivities down to the $\SI{}{fT/\sqrt{\hertz}}$ regime, they usually require cryogenics, heating and/or magnetically shielded environments,  limiting their practicality \cite{boto_meg, schofield_meg_eeg}.

Nitrogen-vacancy (NV) centers in diamond provide a promising alternative \cite{doherty_paper, levine_paper, rondin_paper}. They can be operated at room temperature, offer intrinsic vector capability, and exhibit a large dynamic range (or linear measurement regime/operating range). Recent advances have improved sensitivities to the sub-$\SI{}{pT/\sqrt{\hertz}}$ regime \cite{zhang_sens, barry_sens, graham_sens, sekiguchi_sens, zhang_blueprint, eisenach_sens, wang_sens, fescenko_sens,  xie_sens, gao_sens}. However, sensitivity improvements are often achieved by increasing the coherence time of the NV center, thus narrowing the magnetic resonance linewidth and decreasing the intrinsic dynamic range of the sensor \cite{barry_sens, graham_sens, zhang_blueprint}. General challenges for fluorescence-based sensors are limited collection efficiency and achievable optical contrast between the fluorescence of the spin states due to the spin dynamics and background fluorescence \cite{barry_paper}.

Laser threshold magnetometry (LTM) offers a promising alternative to overcome these  disadvantages and combine high sensitivity with large dynamic range \cite{jeske_ltm, dumeige_ltm, nair_ltm, huck_ltm}. Here, the NV-diamond is placed as an intra-cavity element in the cavity of a laser, and the magnetic field information is read out via the cavity signal. The NV-diamond provides magnetic-field-dependent gain or loss, depending on the wavelength of the laser gain medium. By operating the system close to the threshold, 100\,\% optical contrast between the spin states and strong, coherent signals and thus sensitivities in the few-$\SI{}{fT/\sqrt{\hertz}}$ regime are predicted \cite{jeske_ltm}. This could close the sensitivity gap to SQUIDs and vapor cells, while preserving the advantages of NV magnetometry, including room temperature operation, vector capability, large dynamic range and tiny sensing volumes.

Recent work has demonstrated stimulated emission \cite{jeske_stim_em, nair_amp} and NV-based laser systems \cite{savvin_laser, lindner_paper}, with improvements driven by advances in diamond materials \cite{luo_paper, mironov_superlum, fraczek_spec, capelli_material, luo_abs}. Magnetic-field-dependent stimulated emission \cite{hahl_paper} and two-media LTM schemes at different wavelengths have shown enhanced contrast and sensitivity, particularly near the laser threshold \cite{gottesman_paper, yves_paper}. Further progress has come from cavity-enhanced absorption \cite{jensen_1042nm, chatzidrosos_1042nm, schall_paper}, now observed over a broader wavelength range \cite{schall_paper}, likely linked to phonon sideband absorption of the NV singlet transition \cite{younesi_absorption, kehayias_paper}. Nevertheless, the predicted combination of high contrast with strong measurement signals and thus a significantly improved sensor performance remains the challenge. 

Here, we establish a two-media laser threshold magnetometer at \SI{1042}{\nano\metre} as a new platform for NV-based sensing. By integrating NV centers as an intra-cavity element, our approach achieves \SI{100}{\percent} optical contrast between the spin states and robust signal powers up to \SI{50}{\milli\watt}, while maintaining vectorial operation. This combination enables both improved sensitivity and a large dynamic range, thereby overcoming the trade-off of these sensing-relevant parameters, that have so far limited fluorescence-based NV magnetometers and vapor cell sensors. Our results highlight how this laser-enhanced quantum sensors can pave the way towards a new generation of compact, high-performance magnetic field sensors with broad applicability, from medicine to navigation and resource exploration.

\section*{A nitrogen-vacancy diamond two-media laser system}\label{sec3}
\begin{figure*}[!t]
    \centering
	\includegraphics[scale=1]{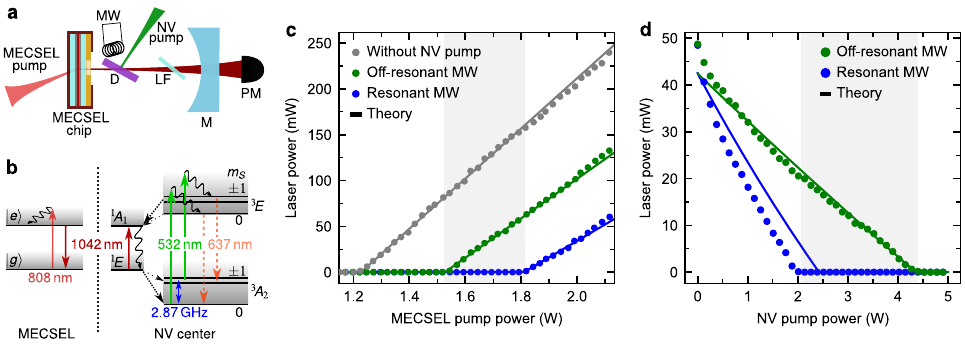}
	\caption{\textbf{Two-media laser system with magnetic-field-dependent laser threshold.} \textbf{a}, Schematic of the linear cavity setup: The cavity is created by the backside mirror of the MECSEL chip and the external cavity mirror (M). The NV-doped diamond (D) is optically pumped with the NV pump and microwave fields are applied with an antenna (MW). A Lyot filter (LF) is used to tune the wavelength to \SI{1042}{\nano\metre}. The cavity signal is detected in transmission with a powermeter (PM). \textbf{b}, Energy levels of the NV center (right) and the MECSEL (left). The NV centers are optically excited to phonon-added sidebands (gray shading) with the green NV pump and can relax back to the ground state via the singlet transition $^1A_1\rightarrow^1E$ (dark red arrow), where they absorb the cavity light emitted from the MECSEL. This NV singlet relaxation path is more probable for the $m_S=\pm 1$ states (thickness of dashed black arrows), leading to a spin-dependent absorption of the cavity field at \SI{1042}{\nano\metre}. The spin state can be probed by scanning a microwave field over the $m_S=0\rightarrow m_S=\pm1$ (blue arrow), where due to the Zeeman effect the resonance frequencies depend on the magnetic field. \textbf{c}, Output power of the laser system as a function of the MECSEL pump power. The laser power varies significantly between off-resonant (green) and resonant (blue) microwaves while the NV centres are optically pumped with \SI{5}{\watt}. This strong signal difference from magnetic resonance can be used for magnetometry. Between the two laser thresholds (gray shading) there is 100\% optical contrast. At \SI{1.8}{\watt} pump power the signal difference is maximal and above \SI{50}{\milli\watt}. The gray curve shows results without optically pumping the NV centers. \textbf{d}, Output power of the laser system as a function of the NV pump power for off-resonant (green) and resonant (blue) microwaves. The MECSEL pump power is constant at \SI{1.3}{\watt}. In both panels the solid lines represent an analytical model with suitably adapted experimental parameters to match the data (see Methods).} 
	\label{fig:figure_1}
\end{figure*}

Motivated by our results of improved spin contrast and sensitivity at higher wavelengths via absorptive magnetometry \cite{schall_paper}, we aim to achieve a boost in sensitivity and dynamic range by realizing the concept of absorptive two-media LTM: a \SI{1042}{\nano\metre} optically active gain medium (a membrane external cavity surface-emitting laser \cite{jetter_MECSEL}, MECSEL) is combined with an NV-doped diamond inside the same cavity, as illustrated in Fig.~\ref{fig:figure_1}a. The NV centers provide a magnetic-field-dependent (or equivalently spin-state-dependent) cavity loss, due to changing population of the absorbing lower singlet state (see Fig.~\ref{fig:figure_1}b) \cite{dumeige_ltm}. We amplify these changes in the cavity field and thus the signal difference between the spin states, i.\,e., the optical contrast, by operating the combined laser system close to the laser threshold \cite{jeske_ltm, dumeige_ltm, gottesman_paper}. Further details concerning the setup and the working principle can be found in Fig.~ \ref{fig:figure_1}a,b and the Methods.

As a first step, we study how the NV centers affect the laser signal via measuring laser power curves, as depicted in Fig.~\ref{fig:figure_1}c. The green and blue curves show how resonant microwaves (blue) applied to the magnetic spin transition shift the laser threshold and change the laser output compared to off-resonant microwaves (green). The gray curve represents the laser power curve without optically pumping the NV centers and thus the diamond losses are reduced to the minimal non-NV-related background losses due to vanishing singlet population. When the microwave field is resonant to the spin transition $m_S=0\rightarrow m_S=\pm 1$ (blue curve), the laser threshold shifts from \SI{1.53}{\watt} to \SI{1.82}{\watt}. Additionally, the slope of the linear curve, i.\,e., the slope efficiency, decreases from \SI{22.0}{\percent} to \SI{18.7}{\percent} indicating a less-efficient laser system. These trends can be well explained by an increased absorption of the cavity field at \SI{1042}{\nano\metre} by the lower singlet state $^1E$ due to increased state population (see Fig.~\ref{fig:figure_1}b). When optically pumping the NV centers, population is also shifted to the singlet states even without a resonant microwave field (see Fig.~\ref{fig:figure_1}b). This also leads to a shift of the laser threshold (\SI{1.23}{\watt} to \SI{1.53}{\watt}) and a decreased slope efficiency (\SI{27.4}{\percent} to \SI{22.0}{\percent}) between the gray and the green curve. This demonstrated spin-dependent absorption of the NV centers results in a broad regime of MECSEL pump powers between the thresholds of the blue and green curve, as indicated by the gray shading in Fig.~\ref{fig:figure_1}c. In this regime the absorption differences are enhanced so strongly that the laser system is switched off by the resonant microwave field. At the point where this effect is strongest at \SI{1.82}{\watt} a laser signal of more than \SI{50}{\milli\watt} is reduced to zero by the resonant microwave field, i.\,e., an optical contrast of 100\,\% is achieved. With these measurements we show a highly promising regime for magnetic field sensing as the signal changes can be equally induced by shifts in the microwave frequency or due to the Zeeman effect by the magnetic field to be sensed (see below).  

As a second step, we study the influence of the NV centers on the laser system in more detail by keeping the MECSEL pump power constant. In Fig.~\ref{fig:figure_1}d the MECSEL output power as a function of the NV pump power is depicted for an off-resonant (green) and a resonant (blue) microwave field. When increasing the NV pump power for both cases the MECSEL output decreases linearly and at an NV pump power of \SI{4.3}{\watt} (with off-resonant microwaves, green), the laser system is turned off due to the NV center absorption. Similarly to the previous paragraph, the cavity losses and thus the slope of the linear decrease can be significantly enhanced by applying a resonant microwave field (blue). There is also a broad regime of NV pump powers for which the MECSEL can be switched off with a resonant microwave field (gray shaded area). These measurements clearly show the difference between a two-media laser system where the NV centers provide gain \cite{lindner_paper, hahl_paper, yves_paper} and where the NV centers provide losses \cite{dumeige_ltm, gottesman_paper} when being optically pumped.

The experiments were very well reproduced by our simulations in which we derived an analytical steady-state solution from a rate model similar to previous work \cite{jeske_ltm, lindner_paper, yves_paper}. The resulting analytical curves are depicted as solid lines in Fig.~\ref{fig:figure_1}c. The analytical model reproduces the experimental data very well. Particularly, for the blue and green curves, the two plot characteristics, threshold value and slope efficiency, change experimentally due to optical pumping and mixing of the NV spin states. Despite two characteristics changing, our model can reproduce these curves by adapting a single parameter, which indicates an accurate description of the two-media laser system by our model. A detailed description of the theoretical model and its adaption to the experiments can be found in the Methods.

\section*{Magnetic field measurements}\label{subsec4}

Next, we measure magnetic fields with the two-media laser system. For this we perform continuous wave optically detected magnetic resonance (ODMR), i.\,e., the NV centers are continuously pumped with the green pump laser and the microwave frequency is scanned over the $m_S$=0$\rightarrow m_S$=$\pm 1$ transition of the ground state $^3A_2$ (see Fig.~\ref{fig:figure_1}b). Simultaneously, the transmitted laser power is measured with a detector. Changes in the resonance frequency and the signal strength can directly be connected to changes in the external magnetic field \cite{levine_paper}. 

\begin{figure*}[!t]
    \centering
    \includegraphics[scale=1]{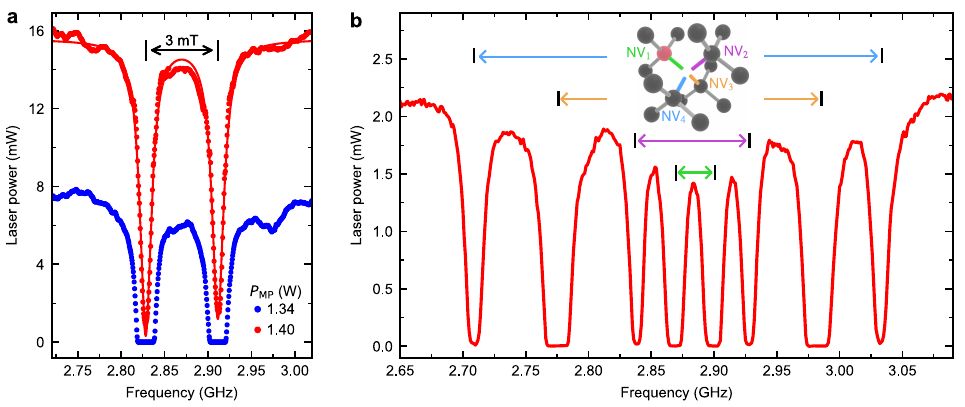}
    \caption{\textbf{Magnetometry with the two-media laser system} \textbf{a}, The blue curve depicts an ODMR measurement, where the laser system is switched off for a resonant microwave field. The red curve shows the same measurement with a slightly increased MECSEL pump power ($P_{\mathrm{MP}}$), demonstrating an ODMR contrast close to \SI{100}{\percent}. The solid line represents a fit with the sum of two Lorentzians. For both measurements, an NV pump power of \SI{5}{\watt} is used and a (100) bias magnetic field is applied with a permanent magnet. \textbf{b}, ODMR with a magnetic field orientation separating all four NV orientations,as indicated with the colored arrows. Vector magnetometry can be performed with \SI{100}{\percent} contrast. The MECSEL pump was set to \SI{1.44}{\watt}. The inset shows a schematic of the NV center in the diamond crystal lattice consisting of carbon atoms (black). We drew the nitrogen atom (red) in one of the four possible orientations (indicated by colors).}  
    \label{fig:figure_2}
\end{figure*}

We first apply a magnetic field in the diamond crystal (100) direction. This separates the $m_S=0\rightarrow m_S=-1$ and the $m_S=0\rightarrow m_S=+1$ transitions and allows to measure magnetic fields via the frequency difference, see Fig.~\ref{fig:figure_2}a. The relative signal difference between resonant and off-resonant signal is called ODMR contrast.  For the blue curve, the MECSEL pump power was chosen such that the MECSEL is turned off for a resonant microwave field due to too high cavity losses, i.\,e., \SI{100}{\percent} contrast is achieved. The red curve in Figure \ref{fig:figure_2}a shows an ODMR measurement with a slightly increased MECSEL pump power. For this case, the MECSEL is not turned off when applying a resonant microwave field. Fitting the data with a sum of Lorentzians (solid line) leads to an ODMR contrast of \SI{97}{\percent} for the deeper resonance and a stronger absolute signal baseline of \SI{16}{\milli\watt} is achieved. We note here that the signal baseline is reduced compared to the power curves in Fig.~\ref{fig:figure_1}c due to the reduced threshold shift of addressing only one spin transition and due to spin-mixing induced by the permanent magnet \cite{hahl_paper, yves_paper}.

To extend the measurement of magnetic fields in only one direction of the field to a configuration where vector magnetometry can be performed, we adjusted the permanent magnet to differently split up all four crystal orientations of the NV centers in the diamond (see inset of Fig.~\ref{fig:figure_2}b). This leads to eight resonances in the ODMR spectrum as depicted in Fig.~\ref{fig:figure_2}b. The separated resonances usually have a weaker contrast than for the case when overlapping the signal from all NV orientations \cite{schloss_vector, barry_sens}. In our laser setup, we simply adjust the MECSEL pump power to achieve again \SI{100}{\percent} contrast for all four orientations, i.\,e., for all components of the magnetic field. The difference in signal strength (different linewidths) for the four NV families is due to the geometric dependence of the optical interaction and the microwave absorption \cite{kehayias_paper, alegre_mw_pol, munzhuber_laser_pol}. This effect cannot be circumvented in our current setup due to the positioning of the diamond at Brewster's angle. Nevertheless, the measurement clearly demonstrates the ability to boost contrast and signal strength for vector magnetometry by two-media LTM. 

\section*{Sensitivity and dynamic range}\label{subsec5}
In magnetometry, the primary goal of a sensor is a large dynamic range in combination with a precise measurement, i.\,e., a good sensitivity \cite{gizzi_paper}. Thus, we investigate how the improved contrast and signal enhancement of two-media LTM has impacted these two quantities.

\begin{figure*}[!t]
    \centering
    \includegraphics[scale=1]{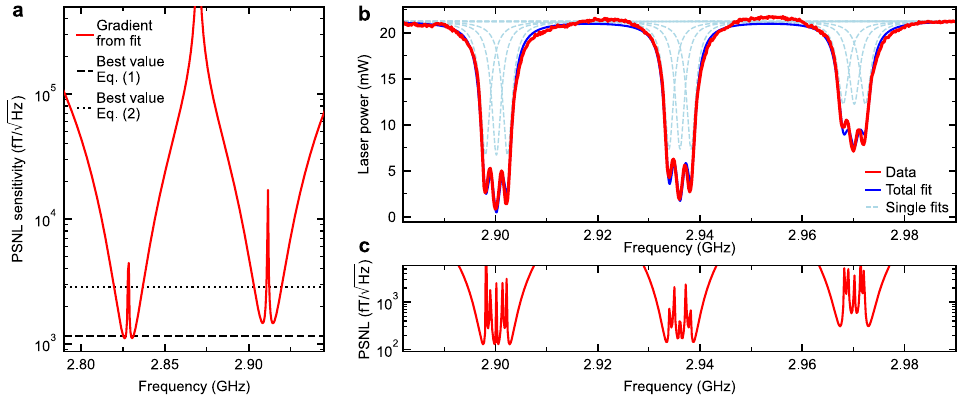}
    \caption{\textbf{Photon-shot-noise-limited sensitivity and its super-linear improvement when approaching 100\,\% contrast.} \textbf{a}, Photon-shot-noise-limited (PSNL) sensitivity of the red ODMR measurement from Fig.~\ref{fig:figure_2}a. The PSNL sensitivity is calculated for each point on the curve via the gradient of the Lorentzian fit (see Methods). The optimal PSNL determined via a generally valid formula and via the commonly used approximation are depicted as dashed and dotted lines, respectively.  An optimum of $\SI{1160}{\femto\tesla\per\sqrt{\hertz}}$ is achieved. Eq.~\ref{eq:sensi_correct} reproduces this value correctly, while the approximation of linear scaling of sensitivity with contrast in Eq.~\ref{eq:sensi_approx} breaks down. \textbf{b} Vector magnetometry measurement with a different diamond sample and slightly adapted experimental parameters (see Methods). Due to the interaction with the neighbouring $^{14}$N atoms, each resonance consists of three hyperfine lines (dashed blue). The NV pump power was set to 3 W. We note here, that the signal from three NV orientations is enough to determine the magnetic field vector \cite{schloss_vector}. \textbf{c} PSNL sensitivity of the ODMR measurement in b, determined via the gradient of the fit (see Methods). An optimum of $\SI{127}{\femto\tesla\per\sqrt{\hertz}}$ is achieved and the sensitivity for all resonances is $<\SI{400}{\femto\tesla\per\sqrt{\hertz}}$.}
    \label{fig:figure_3}
\end{figure*}

The best achievable sensitivity to measure magnetic fields via optical signals and when removing technical noise sources is limited by the fundamental photon shot noise. This photon-shot-noise-limited (PSNL) sensitivity can be calculated from ODMR measurements and varies along the curve. The best PSNL sensitivity $\eta_B$ is calculated as a function of linewidth (full width at half maximum, FWHM) $\Delta\nu$, contrast $C$ and baseline light intensity $I_0$ and was recently derived as a general formula \cite{yves_paper} (details in Methods):

\begin{equation}
\eta_B=\frac{4}{3\sqrt{3}}\frac{h}{g_{\mathrm{e}}\mu_\mathrm{B}}\frac{\Delta\nu}{C\sqrt{I_0}} \frac{\sqrt{(3+S_C^2)^3(3+S_C^2-3C)}}{16S_C}\quad\text{with}\quad S_C=\sqrt{C-1+\sqrt{C^2-5C+4}}.
\label{eq:sensi_correct}
\end{equation}

Here $h$, $g_{\mathrm{e}}$, $\mu_{\mathrm{B}}$ are the Planck constant, electron g-factor and Bohr magneton, respectively. This formula is more commonly used in an approximation for the weak contrast limit $C\ll1$ \cite{levine_paper, barry_paper, hahl_paper, yves_paper, dreau_paper, ruan_paper}, where the best sensitivity scales linearly with contrast 

\begin{equation}
\eta_B \approx \frac{4}{3\sqrt{3}}\frac{h}{g_{\textrm{e}}\mu_\textrm{B}}\frac{\Delta\nu}{C\sqrt{I_0}}. 
\label{eq:sensi_approx}
\end{equation}

For high contrast, the PSNL sensitivity scales super-linearly with the ODMR contrast, i.\,e., improves by more than the factor of contrast improvement \cite{yves_paper}. Thus, we expect an even better PSNL sensitivity when approaching \SI{100}{\percent} ODMR contrast. To investigate this, the red ODMR measurement from Fig.~\ref{fig:figure_2}a was analyzed regarding the PSNL sensitivity. Due to the (100) bias field configuration, an additional geometrical factor of $\sqrt{3}$ needs to be considered when calculating the sensitivity \cite{barry_sens}. The results are depicted in Fig.~\ref{fig:figure_3}a. 

The red curve shows the sensitivity at each point of the curve determined via the gradient of the Lorentzian fit (see Methods). Clearly, the general formula (dashed line) is necessary to determine the correct minimum of the sensitivity curve, while the approximation for small contrast (dotted line) with its linear scaling with contrast underestimates the sensitivity. Thus, we demonstrate a sensitivity which is improved super-linearly with better contrast (details in Methods). 

The dynamic range (or linear measurement regime/operating range) quantifies how much the magnetic signal can vary during a measurement while the sensor output remains linear with the magnetic field \cite{elzenheimer_dynamic_range, gizzi_paper, graham_sens,clevenson_dynamic_range}. For applications in non magnetically-shielded environments, a high dynamic range is preferable because it mitigates the loss of measurement signal or the need for complex regulators to quickly adapt the sensor parameters during sensing \cite{clevenson_dynamic_range, wang_dynamic_range}. Furthermore, a high dynamic range allows for referencing against other sensors or instant analog gradient calculations. Following the definition from \cite{graham_sens}, the dynamic range $\Delta B_{\mathrm{dr}}$ is directly proportional to the ODMR FWHM linewidth $\Delta\nu$ and the electron gyromagnetic ratio $\gamma_\mathrm{e}$ and is given by $\Delta B_{\mathrm{dr}}=\Delta\nu/\gamma_\mathrm{e}$. Particularly, for the typical NV magnetometry approach of frequency-modulated lock-in measurements, this makes sense because a large linewidth corresponds to a large linear regime \cite{graham_sens}. Intrinsically, NV centers can exhibit a large potential dynamic range, but it is often traded to achieve better sensitivity by using narrow-linewidth diamonds leading to $\Delta B_{\mathrm{dr}}<\pm\SI{5}{\micro\tesla}$ for the currently most sensitive NV magnetometers \cite{zhang_sens,barry_sens}. For the red ODMR measurement depicted in Fig.~\ref{fig:figure_2}a, a huge dynamic range of $\Delta B_{\mathrm{dr}}=\pm \SI{280}{\micro\tesla}$, i.\, e., five times larger than the Earth's magnetic field, is achieved. Although it is common to qualify a sensor with the optimal sensitivity in the ODMR curve and take its linewidth as the measure for dynamic range \cite{graham_sens, clevenson_dynamic_range}, Fig.~\ref{fig:figure_2}a shows that optimal sensitivity is achieved for a smaller region than the entire linewidth. We point out that this is generally true for magnetometers based on magnetic resonance and not a special feature of LTM systems.

\begin{figure*}[!tb]
	\centering
	\includegraphics[scale=1]{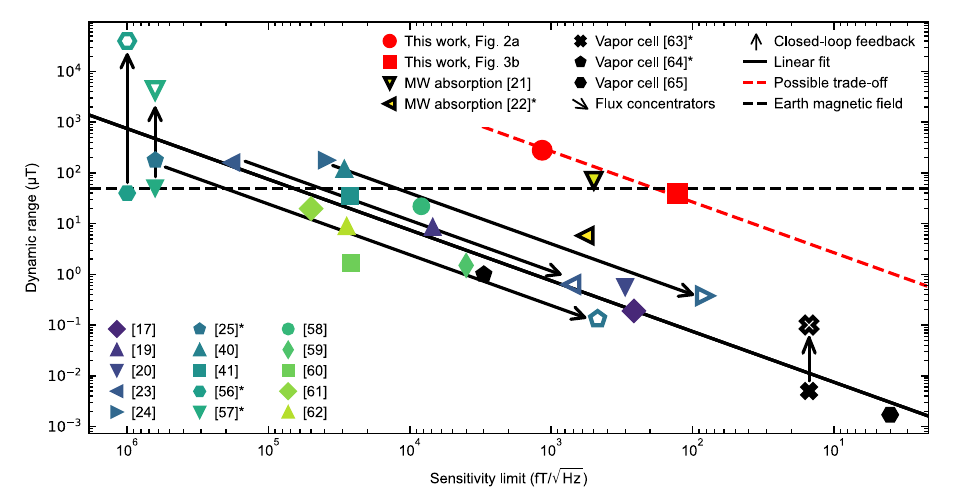}
	\caption{\textbf{Comparison of different magnetometers based on magnetic resonance with the potential for mobile applications.} The desired region is large dynamic range and small sensitivity values (top right corner). NV-based technologies are shown in colors. Black points represent vapor cell magnetometers. Unfilled data points indicate results where the sensitivity was boosted by magnetic flux concentrators, or the dynamic range by closed-loop regulation of either the microwave frequency or the compensating magnetic field. The improvements by these techniques are indicated by arrows. The data points bordered in black represent a different NV magnetometry approach based on cavity-enhanced microwave absorption. The solid line shows a linear fit with a slope of 1, excluding the unfilled and bordered data points, representing the fact that dynamic range can be traded for sensitivity and vice versa. Our results in this work are marked by the red points and show an mean improvement of laser threshold magnetometry by a factor of 520 compared to the solid black line as indicated by the red dashed line. The black dashed line is at \SI{50}{\micro\tesla} and represents the Earth's magnetic field. Data points above this line in principle allow for vector sensing and sensor turning in the Earth's magnetic field. Most of the data was provided directly in the publications or could be calculated from them. For the references marked with a star, no sensitivity limit was available, and thus the measured sensitivity was used instead.}
\label{fig:figure_4}
\end{figure*}

When changing to a diamond sample with higher NV concentration and slightly adapting the experimental parameters (see Methods), a significant sensitivity improvement can be achieved, as illustrated in Fig.~\ref{fig:figure_3}b,c. Here, for all vector components PSNL sensitivites $<$400\,fT/$\sqrt{\textrm{Hz}}$ are achieved, as determined via the gradient of the fit, which now consists of a sum of the three hyperfine lines from the interaction with the $^{14}$N atoms (see Methods). For the most sensitive NV orientation a PSNL sensitivity of 127\,fT/$\sqrt{\textrm{Hz}}$ is demonstrated, marking a new record for a single NV resonance. We note here, that also far away from the threshold PSNL sensitivities $<$400\,fT/$\sqrt{\textrm{Hz}}$ are achievable (see right resonance in Fig.~\ref{fig:figure_3}c). The shown sensitivity improvements compared to Fig.~\ref{fig:figure_3}a are mainly achieved by decreasing the linewidth of the NV resonances and thus trading dynamic range for sensitivity. To further classify the achieved results and this trade-off, different magnetometers based on magnetic resonance and with the potential for mobile applications are compared regarding their dynamic range and sensitivity, as shown in Fig.~\ref{fig:figure_4}.

The combinations of sensitivity limit and dynamic range for NV-based technologies (filled colored points \cite{barry_sens, sekiguchi_sens, zhang_blueprint, dai_sens, zheng_sens, stuerner_sens, patel_sens, kainuma_sens, hahl_paper, gottesman_paper, clevenson_dynamic_range, wang_dynamic_range, xie_sens, gao_sens, fescenko_sens, eisenach_sens, wang_sens}) and vapor cell magnetometers (black data points \cite{greco_opm, quspin_qtfm, ledbetter_opm}) are plotted. We fit a linear trend with a slope of 1 (solid black line), excluding the unfilled and bordered data points. The linear trend represents the fact that dynamic range and sensitivity can each be improved at the cost of the other by the same factor: The dynamic range can be increased via the linewidth, which worsens the signal gradient with the magnetic field and thus the sensitivity and vice versa. Most sensors fit roughly along the linear trend. The aim for a sensor is to achieve high dynamic range and good sensitivities (upper right corner of Fig.~\ref{fig:figure_4}) \cite{gizzi_paper}. Our results in this work (red data points) clearly show a strong improvement from the linear behavior towards the desired regime. The improvement in the ratio of dynamic range and the PSNL sensitivity with our setup is up to 590 compared to the linear fit. This quantifies the strong performance boost achieved by the laser cavity feedback, which leads to \SI{100}{\percent} contrast and strong laser signals. Vertical improvements from the linear behavior can be achieved by using a closed-loop feedback \cite{clevenson_dynamic_range, wang_dynamic_range, greco_opm}, where the sensors parameters are adjusted to hold the resonance criterion during measurements. We have not implemented closed-loop feedback, but it provides a path for further improvement. However, enlarging the dynamic range via closed-loop feedback places restrictions on how fast magnetic field changes or movements in the magnetic field can happen without loosing the feedback lock as well as limiting the signal bandwidth due to the required modulation frequency. Thus, sensors without feedback extensions of the dynamic range also have advantages. Another strong deviation from the linear behavior is marked by the black-bordered data points \cite{eisenach_sens, wang_sens}. There, a different NV-based magnetometry concept of cavity-enhanced microwave absorption was used and \SI{100}{\percent} absorption contrast was demonstrated for scalar magnetometry, leading to a notable sensitivity improvement. But when combining this approach with vector magnetometry significantly worse contrast ($C\leq\SI{40}{\percent}$) and thus sensitivity is achievable \cite{wilcox_vector_mw}. The sensitivity of NV magnetometers can also be strongly improved when magnetic flux concentrators are used \cite{xie_sens, gao_sens, fescenko_sens}. However, these effectively reduce the relative ODMR linewidth in units of magnetic field and thus the dynamic range (indicated by the arrows parallel to the linear fit). Magnetic flux concentrators could be used to further shift our sensor performance from its current position parallel to the linear trend (red dashed line). By comparing our sensor to other state of the art NV-based and vapor cell sensors in Fig.~\ref{fig:figure_4}, we clearly see the improved performance and the potential of LTM to boost the sensitivity and/or the dynamic range of NV magnetometers significantly, while maintaining their advantages. 

\section*{Discussion and outlook}
In this work, we demonstrate how integrating the quantum sensing material NV diamond inside a laser cavity and reading out via the laser signal can boost quantum sensing. We show an improvement in terms of intrinsic dynamic range and/or sensitivity by a factor of 780 up to 590 compared to the commonly used fluorescence-based readout and atomic vapor cell magnetometers. In absolute values, we achieve a dynamic range of $\pm$\SI{280}{\micro\tesla} (set by the ODMR linewidth) in combination with a PSNL sensitivity of $\SI{1160}{fT/\sqrt{\hertz}}$, without the use of magnetic flux concentrators. When changing to a different diamond sample and adjusting the experimental parameters, we show in vector configuration a PSNL sensitivity of $<$400\,fT/$\sqrt{\textrm{Hz}}$ in combination with a dynamic range of $\pm$\SI{40}{\micro\tesla}, demonstrating the possible linear trade-off between these two quantities.  We show \SI{100}{\percent} optical contrast, i.\,e., we are able to entirely switch off the sensor laser system with the magnetic resonance. At the same time we achieve strong signal intensities in a collimated beam between \SI{2}{\milli\watt} and \SI{50}{\milli\watt}, while still allowing for vector magnetometry with \SI{100}{\percent} contrast. This enables multiple breakthroughs in the field of NV magnetometry:  an improvement factor by more than two orders of magnitude and the first super-linear improvement of PSNL sensitivity with contrast. The contrast is improved well beyond the low-excitation limit of fluorescence-based NV sensors of about \SI{22}{\percent} \cite{hahl_paper}, which experimentally is typically much smaller \cite{barry_sens, graham_sens, sekiguchi_sens}. 

These improvements are enabled by the realization of the concept of laser threshold magnetometry with two intra-cavity media: a MECSEL medium creates gain at \SI{1042}{\nano\metre} while the NV-doped diamond acts as a magnetic-field-dependent (or equivalently spin-dependent) absorber at this wavelength.Our results were achieved in a simple continuous wave ODMR measurement and the achieved PSNL sensitivity is  better than the currently best achieved PSNL sensitivity for NV magnetometry \cite{barry_sens}. There, in total 15 techniques, such as  pulse sequences and P1 driving were used, which boost the PSNL sensitivity as well as remove technical noise. None of these techniques were applied in our setup yet, leaving a lot of room for future improvements to boost our magnetometry approach and to remove technical noise sources. In addition, significant improvements in sensitivity could be achieved by combining our sensors with magnetic flux concentrators, which have been shown to enable a more than 1000-fold improvement at the cost of dynamic range and vector capabilities \cite{gao_sens}. This could in principle enable sub-$\SI{}{fT/\sqrt{\hertz}}$ sensitivities, while maintaining the factor 520 mean advantage from Fig.~\ref{fig:figure_4} and the general benefits of NV-based magnetometry. Lastly, improvements are expected by exploring and optimizing the large parameter space of other laser materials, diamond properties, and cavity components as well as improving homogeneity in the microwave delivery and bias field of the sensor.

The PSNL sensitivity is the limit of sensitivity which can be achieved in a setup when all technical noise sources are eliminated. In NV-based magnetometry and vapor cell magnetometers, real sensitivities within a factor of 3 of the shot noise limit are commonly achieved \cite{barry_sens, sekiguchi_sens, chatzidrosos_1042nm, younesi_1042nm, mitchell_snl_opm}. Our experimental setup was designed as a proof-of-concept and the comparison in Fig.~\ref{fig:figure_4} shows the significant possible advantage of laser threshold magnetometry. Real sensitivities approaching our shot noise limit can be achieved with a redesign towards a highly-stable laser system, lock-in detection and balanced detection for common-mode noise rejection. An additional challenge in our system is the cavity stabilization and the fact that two pump lasers are used and pump laser noise from both lasers needs to be removed via advanced balancing techniques such as digital balancing. However, comparable semiconductor-based laser systems have been shown to approach the photon shot noise in the relevant frequency regime \cite{hastie_psnl} and a lot of research is going on to miniaturize and stabilize such systems via monolithic integration \cite{hastie_monolithic, 21s_monolithic}. Therefore, we see a clear path to laser threshold vector magnetometers with real sensitivities and improvement factors as determined in this work. 

The shown possible improvements in sensor performance could make NV centers in diamond increasingly more relevant and enable a new generation of quantum sensors, especially for medical applications, boosting dynamic range and/or sensitivity required for measuring physiological and neurological signals. The improved dynamic range furthermore provides a perspective to create sensors and gradiometers which can be operated, worn and turned within the Earth's magnetic field without magnetic shielding. This could lead to new technologies that are no longer limited to expensive magnetically shielded rooms. Due to the intrinsic nature of the dynamic range, these new sensors could even be operated without the need for feedback loops which adapt the sensor conditions while being moved on the Earth's background field. Thus, they could get rid of speed restrictions, bandwidth limitations and instabilities due to such feedback lockings. 

Our results show the principle of using laser-enhancement for boosting performance in quantum sensors and could be transferred to other emitters in diamond, SiC or other optical emitter materials as well as sensors based on optical absorption such as atomic vapor cells. In general, the results provide a new perspective on coupling an ensemble of controllable quantum systems to a laser output and enable improved readout and performance through the enhancement of the laser cavity.

\clearpage
\thispagestyle{plain}

\section*{Methods}\label{sec11}
\subsection*{Experimental setup and methodology}
The two-media laser system used in this work is schematically shown in Fig.~\ref{fig:figure_1}a . The optical gain medium is a membrane external cavity surface-emitting laser chip (MECSEL) \cite{jetter_MECSEL}, which has a broad gain spectrum with a maximum at \SI{1075}{\nano\metre} in an empty cavity. The gain chip was grown on a 4-inch GaAs substrate and consists of 11 InGaAs quantum wells. The membrane is sandwiched between two SiC heat spreaders with a thickness of \SI{500}{\micro\metre}. One side of this stack has a highly-reflective coating for the MECSEL bandwidth and anti-reflective coating for the pump wavelength at \SI{808}{\nano\metre} and thus represents the first plane cavity mirror. The other side of the stack has an anti-reflective coating for the MECSEL bandwidth. Without any frequency filtering inside the cavity, the MECSEL runs multi-mode and exhibits a broad emission spectrum of approximately \SI{6}{\nano\metre} with the center at \SI{1075}{\nano\metre}. The MECSEL gain spectrum is broad, making it possible to tune the wavelength down to \SI{1042}{\nano\metre} with an intra-cavity birefringent filter (see below). The gain chip is optically pumped from the backside with a fiber-coupled diode laser at \SI{808}{\nano\metre} ($d_{\textrm{core}}=\SI{105}{\micro\metre}$, $\textrm{NA}=0.22$), which we call ``MECSEL pump''. The pump is collimated (Thorlabs F810SMA-780) and focused on the chip with a lens ($f=\SI{50}{\milli\metre}$) resulting in a pump spot diameter of around \SI{150}{\micro\metre} in the focus. The second cavity mirror (M) is concave ($R=\SI{95.1}{\percent}$, ROC=\SI{100}{\milli\metre}) at a distance of $L_{\mathrm{cav}}\approx\SI{98.5}{\milli\metre}$. This cavity geometry results in a cavity mode diameter of around \SI{140}{\micro\metre} on the gain chip leading to good mode overlap between the MECSEL pump and the cavity mode. 

The diamond (D) is placed \SI{28}{\milli\metre} away from the gain chip. Due to the cavity geometry, the cavity mode beam diameter at the diamond position is \SI{280}{\micro\metre}. The diamond is oriented at Brewster's angle, leading to reduced cavity losses for a p-polarized MECSEL mode. Two slightly different, (100) oriented type 1b high-pressure-high-temperature (HPHT) diamond samples from the same batch were used in this work. Both diamonds have a geometry (\num{3}$\times$\num{3}) \SI{}{\milli\metre\squared} and a comparable thickness of $\approx$\SI{300}{\micro\metre}. The yellow starting material was commercially obtained from Element Six. The diamond used in Fig.~\ref{fig:figure_1} and Fig.~ \ref{fig:figure_2} was pretreated with a low-pressure-high-temperature (LPHT) annealing at \SI{1800}{\degree C} in a hydrogen atmosphere. This process leads to a decrease in the diamond absorption \cite{luo_abs, hahl_phd}. After this treatment, the diamond was electron irradiated with a fluence of \SI{1e18}{\per\centi\metre} and an electron energy of \SI{2}{\mega\electronvolt}. The diamond was then annealed for 2 hours at \SI{1000}{\degree C} to create nitrogen-vacancy (NV) centers. A high NV$^-$ concentration of \SI{1.91}{ppm} was achieved, as was determined with the photoluminescence method from our previous work \cite{luo_paper}. The diamond used in Fig.~\ref{fig:figure_3} was electron irradiated with a higher dose of \SI{2e18}{\per\centi\metre}, yielding a higher NV$^-$ concentration of \SI{3.47}{ppm}. Due to the much higher NV concentration less NV pump power and microwave power is necessary to obtain a strong NV signal. Therefore, the linewidth, i.\,e. the dynamic range can be traded to achieve a better photon-shot-noise-limited sensitivity, as demonstrated in Fig.~\ref{fig:figure_4}. For both diamonds the coherence times are comparable ($T_2^*\approx\SI{200}{\nano\second}$), as was determined in a confocal photoluminescence setup with a Ramsey sequence.

The NV centers are optically pumped with a \SI{532}{\nano\metre} laser, which we call ``NV pump''. It is focused on the diamond with a lens ($f=\SI{125}{\milli\metre}$) and adjusted to achieve good mode-matching with the cavity mode. The NV pump optically excites the NV centers from the triplet ground state $^3A_2$ to the excited state $^3E$ (see green arrow in Fig.~\ref{fig:figure_1}b). From there, two relaxation paths back to the ground state exist. On the one hand, the NV center can fall back to the ground state under emission of a photon in the wavelength range from \SI{637}{\nano\metre} up to \SI{850}{\nano\metre} (orange dashed arrows) \cite{levine_paper}. On the other hand the NV center can undergo intersystem crossing (ISC, dashed black arrows) to the excited singlet state $^1A_1$, followed by the singlet transition to the long-lived $^1E$ and back to the ground state via ISC. The singlet transition (dark red arrow) exhibits a zero phonon line (ZPL) of \SI{1042}{\nano\metre} and phonon broadening seems to lead to a broadband absorption features of the lower singlet state $^1E$ \cite{schall_paper, kehayias_paper, younesi_absorption} despite very weak phonon sideband emission compared to the singlet ZPL emission \cite{rogers_paper}. The singlet transition is far less radiative than the triplet transition due to increased phonon transitions. Furthermore, the ISC is more probable for the $m_S=\pm1$ states (thickness of dashed arrows), making it possible to initialize to $m_S=0$ and to readout the spin state via a change in fluorescence after or during optical excitation \cite{levine_paper}. Alternatively, the spin state can be determined via a change in absorption of a probe beam at \SI{1042}{\nano\metre} \cite{jensen_1042nm, chatzidrosos_1042nm, acosta_1042nm, acosta_singlet, dumeige_1042nm, younesi_1042nm}. The absorption is increased for the $m_S=\pm1$ states due to a favorable ISC and thus a higher population of the absorbing lower singlet state $^1E$. The spin-dependence of spontaneous emission and of absorption at \SI{1042}{\nano\metre} is strongly enhanced by the long lifetime of the metastable lower singlet state $^1E$ ($>$\SI{200}{\nano\second} \cite{acosta_1042nm}).

A microwave field (MW) around \SI{2.87}{\giga\hertz} is used to coherently drive the transition between $m_S=0$ and $m_S=\pm 1$ in the triplet ground state $^3A_2$ (blue arrow in Fig.~\ref{fig:figure_1}b). Due to the Zeeman effect, the resonance frequencies of these spin transitions depend on the external magnetic field \cite{levine_paper}. The microwave setup is the same as in previous work \cite{yves_paper, schall_paper}.

A Lyot filter (LF) with a thickness of \SI{3}{\milli\metre} is used to tune the wavelength of the MECSEL down to the zero phonon line (ZPL) of the singlet transition at \SI{1042}{\nano\metre}. The filter is made of crystal quartz, which is birefringent, leading to a wavelength-dependent phase difference $\delta_{\lambda}$ between the ordinary and extraordinary beam \cite{milonni_laser,lyot_evans,lyot_bloom}. Only when the phase difference satisfies $\delta_{\lambda}=2\pi$ does the Lyot filter preserve the polarization of the incident light, thereby minimizing cavity losses. By placing the filter at Brewster's angle, only p-polarized modes will be allowed for the MECSEL due to decreased cavity losses. By rotating the optical axis of the Lyot filter the refractive index for the ordinary and extraordinary and thus the phase difference $\delta_{\lambda}$ are changed, leading to a change in the MECSEL wavelength.

The MECSEL signal is measured in transmission with a powermeter (PM). A \SI{1000}{\nano\metre} longpass filter is used to block the light from the two pump lasers. The transmitted cavity signal is also fiber coupled to a grating spectrometer to resolve the spectrum of the two-media laser system. Usually, there is mode competition between different MECSEL modes with a distinct spectral distance of \SI{0.4}{\nano\metre} due to the intra-cavity etalon introduced by the \SI{500}{\micro\metre} thick SiC heat spreaders. We can monitor these modes in the spectrometer and the laser system can be adjusted such, that only one of those heat spreader modes turns on. This adjustment is strongly dependent on the operation point of the laser system, e.\,g., on the MECSEL pump power and the intra-cavity field. This leads to multi-mode operation, when the system is operated far from the threshold, as can be seen in deviations from the linear laser behavior in  Fig.~\ref{fig:figure_1}c,d.

\subsection*{Theoretical model}
The theoretical model to describe the two-media laser system analytically is based on methods previously used \cite{jeske_ltm, lindner_paper, yves_paper}. In the following the analytical model is described and afterwards the adaptation of the parameters to match the experimental data from the main text is discussed. 

\begin{figure}[!h]
	\centering
	\includegraphics[scale=1]{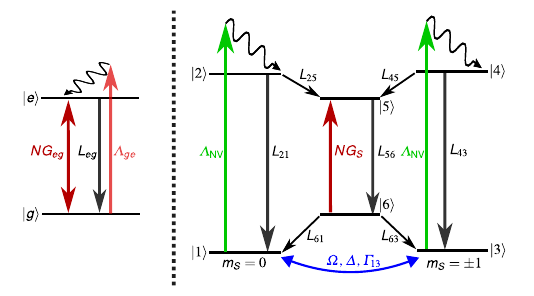}
    \renewcommand{\figurename}{}
    \renewcommand{\thefigure}{Extended Data Fig. 1}
	\caption{\textbf{Energy levels and relevant transitions for the rate model describing the two-media laser system.} The NV center is modeled as a six-state system in combination with an optical excitation (green, $\Lambda_{\mathrm{NV}}$), spontaneous emission (black, $L_{21}=L_{43}$) and a coherent microwave drive (blue, $\Omega, \Delta, \Gamma_{13}$).  The MECSEL is modeled as a two level system, which is optically pumped (red, $\Lambda_{\mathrm{ge}}$), has a loss rate (black, $L_{\mathrm{eg}}$) and a gain to the cavity field (dark red, $G_{\mathrm{eg}}$). The NV center absorbs photons of the cavity field $N$ with an absorption strength $G_{S}$ (dark red), leading to a coupling of the spin dynamics of the NV center and the cavity field.}
\label{fig:figure_S1}
\end{figure}

The model is based on finding an analytical steady-state solution to the equations of motion (EOMs) for both systems: the NV centers in diamond and the MECSEL. The two systems are coupled via the total number of photons inside the cavity. The relevant energy states and the nomenclature for the description are depicted in \ref{fig:figure_S1}. By comparing this schematic with the level structure in Fig.~\ref{fig:figure_1}b, one can easily assign the energy states and the optical transitions, which are relevant for the model. For simplicity, the MECSEL is assumed as a two-level emitter. The strength of the NV and the MECSEL pump are given by the pumping rate $\Lambda_{\mathrm{NV}}$ and $\Lambda_{\mathrm{ge}}$, respectively. The $L_i$ are the transition rates between different energy states. The NV ground state spin states are modeled coherently, i.\,e., with on- and off-diagonal density matrix elements, while all other transitions are modeled as rate equations. The coherent microwave coupling between $m_S=0$ and $m_S=\pm1$, i.\,e., the coupling between $\ket{1}$ and $\ket{3}$, is described as a Rabi oscillation with Rabi frequency $\Omega$, detuning to the resonance $\Delta$ and dephasing $\Gamma_{\mathrm{13}}=1/T_2^*$. The coupling of the NV centers and the MECSEL to the cavity field is given by the cavity coupling strengths $G_{S}$ and $G_{eg}$, respectively. Due to the short (long) lifetime of the upper (lower) singlet state $\ket{5}$ ($\ket{6}$) \cite{acosta_singlet} and the non-radiative decay pathway in the singlet transition, the NV center's cavity effect is strongly dominated by absorption and stimulated emission is negligible. Taking the contribution of stimulated emission into account led to the same results while having more complex analytical expressions. Thus, the NV center's cavity coupling $G_{S}$ is modeled as a purely absorptive coupling. The strength of the cavity field is described by the number $N$ of cavity photons. 

Following the described nomenclature and the same procedure as in \cite{jeske_ltm, lindner_paper, yves_paper}, where the density matrix elements $\rho_{11}, \rho_{22}, \dots \rho_{66}, \rho_{gg}, \rho_{ee}$ are populations of the corresponding states $\ket{j}$ and the off-diagonal density matrix elements $\rho_{13}, \rho_{31}$ are the coherences of the corresponding states $\ket{1}$ and $\ket{3}$, the EOMs of the two systems are given by the following master equation

\begin{equation}
	\dot{\rho}_{13}=(\mathrm{i}\Delta-\Gamma_{13}-\Lambda_{\mathrm{NV}})\rho_{13}+\mathrm{i}\Omega(\rho_{11}-\rho_{33}),
	\label{eq:rho_13}
\end{equation}
\begin{equation}
	\dot{\rho}_{31}=-(\mathrm{i}\Delta+\Gamma_{13}+\Lambda_{\mathrm{NV}})\rho_{31}+\mathrm{i}\Omega(\rho_{33}-\rho_{11}),
	\label{eq:rho_31}
\end{equation}
\begin{equation}
	\dot{\rho}_{11}=\mathrm{i}\Omega(\rho_{13}-\rho_{31})-\Lambda_{\mathrm{NV}}\rho_{11}+L_{21}\rho_{22}+L_{61}\rho_{66},
	\label{eq:rho_11}
\end{equation}
\begin{equation}
	\dot{\rho}_{22}=\Lambda_{\mathrm{NV}}\rho_{11}-(L_{21}+L_{25})\rho_{22},
	\label{eq:rho_22}
\end{equation}
\begin{equation}
	\dot{\rho}_{33}=\mathrm{i}\Omega(\rho_{31}-\rho_{13})-\Lambda_{\mathrm{NV}}\rho_{33}+L_{43}\rho_{44}+L_{63}\rho_{66},
	\label{eq:rho_33}
\end{equation}
\begin{equation}
	\dot{\rho}_{44}=\Lambda_{\mathrm{NV}}\rho_{33}-(L_{43}+L_{45})\rho_{44},
	\label{eq:rho_44}
\end{equation}
\begin{equation}
	\dot{\rho}_{55}=L_{25}\rho_{22}+L_{45}\rho_{44}-L_{56}\rho_{55}+G_S\rho_{66}N/N_{\mathrm{NV}},
	\label{eq:rho_55}
\end{equation}
\begin{equation}
	\dot{\rho}_{66}=L_{56}\rho_{55}-G_S\rho_{66}N/N_{\mathrm{NV}}-(L_{61}+L_{63})\rho_{66},
	\label{eq:rho_66}
\end{equation}
\begin{equation}
	\dot{\rho}_{gg}=-(\Lambda_{ge}+G_{eg}N/N_{\mathrm{2M}})\rho_{gg}+(L_{eg}+G_{eg}N/N_{\mathrm{2M}})\rho_{ee},
	\label{eq:rho_gg}
\end{equation}
\begin{equation}
	\dot{\rho}_{ee}=(\Lambda_{ge}+G_{eg}N/N_{\mathrm{2M}})\rho_{gg}-(L_{eg}+G_{eg}N/N_{\mathrm{2M}})\rho_{ee}.
	\label{eq:rho_ee}
\end{equation}

Here, the contribution from the MECSEL and the NV centers are taken into account by normalizing the number of cavity photons $N$ to the number of MECSEL emitters $N_{\mathrm{2M}}$ and NV centers $N_{\mathrm{NV}}$, respectively. For simplicity, the two quantities are assumed as equal and the strength of the contribution from both media is tuned via changing the pumping rates $\Lambda_{ge},\ \Lambda_{\mathrm{NV}}$ and cavity couplings $G_{eg},\ G_{S}$. The NV pumping rate is connected to the absorption cross section $\sigma_{532}$ at \SI{532}{\nano\metre} via $\Lambda_{\mathrm{NV}}=\sigma_{532}I_{\mathrm{NV}}/E_{\mathrm{ph}}$, where $I_{\mathrm{NV}}$ is  the pump intensity and $E_{\mathrm{ph}}$ is the photon energy. The MECSEL pumping rate is chosen two orders of magnitude larger. The two systems are coupled via the EOM for the time-dependent number of cavity photons $N$, taken the total loss rate of the cavity $\kappa$ into account:

\begin{equation}
	\dot{N}=-G_SN\rho_{66}+G_{eg}N(\rho_{ee}-\rho_{gg})-\kappa N.
	\label{eq:N}
\end{equation}

To solve the system the same procedure as in \cite{jeske_ltm, lindner_paper, yves_paper} is applied: We solve only for the steady-state solution $\dot\rho =0, \dot N=0$. In the steady-state solution all variables are time-independent constants and thus Eq.~\ref{eq:rho_13} to Eq.~\ref{eq:rho_ee} can be solved independently, treating $N$ as an unknown but constant parameter. We write the set of equations for both systems in (superoperator) matrix formalism. Then we solve for the kernel/nullspace of the matrix to get a solution for the density matrix elements dependent on all parameters, including the unknown $N$. Then, these solutions are inserted in Eq.~\ref{eq:N} and the steady-state solution ($\dot{N}=0$) is solved. We obtain an analytical solution using Mathematica. The parameters which we used are literature values or are determined from our experimental setup and are summarized in \ref{tab:parameters}. 

The concentration of NV centers was measured via a calibrated photoluminescence brightness measurement with the procedure described in \cite{luo_paper} and then multiplied with the overlap of the cavity mode volume with the diamond plate to obtain the number of NV centers $N_{\mathrm{NV}}$ contributing to the signal. The spin state decoherence rate $\Gamma_{13}=1/T_2^*$ was determined by measuring the $T_2^*$ coherence time in a confocal photoluminescence setup with a Ramsey sequence. The cavity loss rate $\kappa$ was estimated by measuring the slope efficiency of the laser system for varying reflectivity of the cavity mirror as described in \cite{jetter_MECSEL, kuznetsov_vecsel}.  
  
To test whether our theoretical model can describe the experiment, the data from Fig.~\ref{fig:figure_1}c is fitted with the analytical expression. We obtain the expression by taking the analytical solution for the number of cavity photons $N$, interpret it as a function of the MECSEL pump rate $\Lambda_{ge}$ and then convert to laser powers with $P_{\mathrm{out}}=E_{\mathrm{ph}} \kappa_{\mathrm{mirror}} \, N(\Lambda_{ge})$. Here, $E_{\mathrm{ph}}$ is the photon energy, $\kappa_{\mathrm{mirror}}=\SI{75}{\mega\hertz}$ is the cavity loss rate due to the outcoupling mirror and $N(\Lambda_{ge})$ is the photon number as a function of the MECSEL pump rate. The MECSEL pump power is $P_{\mathrm{MP}} \propto \Lambda_{ge}$ with the assumed proportionality factor given in \ref{tab:parameters}. We then adapt the following parameters to obtain a good match between our model and the experimental curves by fitting the given analytical expression to the data. To accurately describe the behavior around the threshold, only the first 10 data points above the threshold were used for this adaption. First, the gray data in Fig.~\ref{fig:figure_1}c is fitted to obtain the two-level-emitter's (MECSEL) spontaneous emission rate $L_{eg}$ and the cavity coupling $G_{eg}$. The results are given in \ref{tab:fit_results}. With this we have obtained the MECSEL parameters which we later use to reproduce the other two curves in the same figure.

An alternative simpler way to fit the MECSEL parameters confirms the same result: Without optically pumping the NV centers, the first term in Eq.~\ref{eq:N} can be neglected since the populations are $\rho_{55}=\rho_{66}=0$. When assuming detailed balance for the two-level system of the MECSEL, the steady-state solution for Eq.~\ref{eq:N} is given as 

\begin{equation}
N=\frac{(G_{eg}-\kappa)N_{\mathrm{2M}}}{2\kappa G_{eg}}\Lambda_{ge}-\frac{(G_{eg}+\kappa)N_{\mathrm{2M}}}{2\kappa G_{eg}} L_{eg}.
\label{eq:N_easy}
\end{equation} 

This describes a linear curve of the cavity photons $N$ in respect to the pumping rate $\Lambda_{ge}$. Fitting this comparably easy expression to the gray data from Fig.~\ref{fig:figure_1}c leads to the same parameters for the MECSEL (see \ref{tab:fit_results}). This validates that the complex analytical expression obtained by the procedure explained above describes the case without optical pumping accurately. 

To test the model in regard to the contribution of the NV centers, the green data from Fig.~\ref{fig:figure_1}c, where the green NV pump laser is turned on, is fitted with the analytical expression. The MECSEL parameters determined by the previous fitting are inserted and the detuning is chosen to be far away from the resonance ($\Delta=\SI{0.87}{\giga\hertz}$, as was used in the experiments). Thus, the only free parameter for the fitting is $G_S$, which describes the strength of cavity field absorption by the NV centers, i.\,e., the coupling of the NV centers to the cavity field. The obtained value for $G_S$, which makes the analytical solution fit the obtained data best, is given in \ref{tab:fit_results}. We point out that by turning on the green NV pump laser (and thus the NV absorption) two parameters in the plot are changed: the laser threshold and the slope of the curve, i.\,e., the slope efficiency of the laser. With the variation of a single model parameter $G_S$ to a complicated analytical solution we were able to reproduce both parameter changes, indicating that the model is a good description of the physics in the experimental data.

Next, we regard the blue curve in Fig.~\ref{fig:figure_1}c, where additionally the microwave drive is turned resonant and thus Rabi driving mixes the ground state spin states. We use again one free parameter in the analytical solution: the Rabi driving strength $\Omega$, to now reproduce the blue data with our model. The MECSEL parameters and the strength of the NV absorption obtained before are inserted in the model. The obtained value for $\Omega$, which makes the analytical solution fit the obtained data best, is given in \ref{tab:fit_results}. Again, two experimental parameter changes (threshold and slope) are reproduced by adapting one model parameter.

The fitting for the data in Fig.~\ref{fig:figure_1}d was performed slightly different. The green curve is matched to the data by adapting the MECSEL parameters ($L_{eg}=\SI{5.1}{\mega\hertz}$, $G_{eg}=\SI{354}{\mega\hertz}$). We note here that these adaptations are justified due to slightly different experimental conditions, like cavity adjustments, NV pump overlap and antenna positioning compared to Fig.\ref{fig:figure_1}c. The blue curve was obtained from the same MECSEL parameters as the green curve and by inserting the values for $G_S$ and $\Omega$ from \ref{tab:fit_results}. The resulting threshold shift (without adapting any parameter) reproduces the experimental threshold shift within reasonable bounds.

\subsection*{Photon-shot-noise-limited sensitivity}
The continuous wave (cw) sensitivity $\eta_B=\sigma_B\sqrt{T}$ is a measure of the precision of a magnetometer which is independent of the measurement time $T$. The standard deviation of a magnetic field $\sigma_B$ can be rewritten using error propagation as $\sigma_B=|\partial B/\partial I|\sigma_I$. Here, $I$ is the photon rate in the signal of optically detected magnetic resonance (ODMR) experiments. The standard deviation $\sigma_I$ in the shot noise regime can be rewritten using Poissonian statistics as $\sigma_I=\sqrt{I}/\sqrt{T}$ \cite{dreau_paper}. This leads to the general definition of the photon-shot-noise-limited (PSNL) sensitivity as given by 

%\begin{equation}
%\eta(\nu)=\frac{dB}{dI} \sqrt{I(\nu)}=\frac{2 \pi \hbar}{g_e %\mu_B}\left(\frac{dI}{d\nu}\right)^{-1}\sqrt{I(\nu)}
%\end{equation}
\begin{equation}
\eta_B(\nu)=\left|\frac{\partial B}{\partial\nu}\frac{\partial\nu}{\partial I}\right| \sigma_I \sqrt{T} =\frac{h}{g_{\mathrm{e}}\mu_\mathrm{B}}\left|\frac{\partial I}{\partial\nu}\right|^{-1}\sqrt{I(\nu)}.
\label{eq:psnl_general}
\end{equation}

Here $h$, $g_{\mathrm{e}}$, $\mu_{\mathrm{B}}$ are the Planck constant, electron g-factor and Bohr magneton, respectively. The ODMR curve of the photon rate as a function of the microwave frequency is normally Lorentzian-shaped,

\begin{equation}
I(\nu)=I_0\left(1-\frac{C\Delta\nu^2}{\Delta\nu^2+4\nu^2}\right),
\end{equation} 

where $\nu$ is the frequency shift from the resonance, $\Delta\nu$ is the linewidth
(full width at half maximum, FWHM), $C$ is the contrast and $I_0$ is the baseline of the Lorentzian \cite{dreau_paper}. The photon rate is connected to the detected power $P$ and the energy per photon $E_{\mathrm{ph}}$ via $I(\nu)=P(\nu)/E_{\mathrm{ph}}$. At this point, usually the assumption is made that the operating point of the best sensitivity in the ODMR curve $\eta_B=\eta_B(\nu_{\mathrm{ip}})$ is achieved at the point of maximal slope in the Lorentzian curve, i.\,e., the inflection point $\nu_{\mathrm{ip}}=\Delta\nu/(2\sqrt{3})$ and that the signal strength at this point can be approximated by the baseline power $I_0$ of the ODMR curve. These assumptions are valid only for small contrast and lead to the expression 

\begin{equation}
\eta_B\approx \frac{4}{3\sqrt{3}}\frac{h}{g_{\mathrm{e}}\mu_\mathrm{B}}\frac{\Delta\nu}{C\sqrt{I_0}}
\label{eq:psnl_old}
\end{equation}

for the photon-shot-noise-limited (PSNL) sensitivity of a cw magnetometer \cite{dreau_paper}. From this formula it is often assumed, that the sensitivity scales linearly with enhanced contrast, however, this is only valid in the small contrast regime. In laser threshold magnetometry (LTM), where strong contrasts are achieved, this formula is not valid. The signal strength at the inflection point is $I(\nu_{\mathrm{ip}})=I_0(1-3C/4)$. Thus, the assumption $I(\nu_{\mathrm{ip}})\approx I_0$ is only valid for small contrast. Furthermore, the point of optimal sensitivity is shifted for larger contrast from the inflection point closer to the minimum of the ODMR curve. The fully correct formula for the PSNL sensitivity without the small contrast approximation is determined by not fixing the operating point to the inflection point and not fixing the signal to the baseline signal. The formula is derived in detail in \cite{yves_paper} and can be given as a correction factor to the approximated formula 

\begin{equation}
\eta_{B}= \frac{4}{3\sqrt{3}}\frac{h}{g_{\mathrm{e}}\mu_\mathrm{B}}\frac{\Delta\nu}{C\sqrt{I_0}} \,\, \frac{\sqrt{(3+S_C^2)^3(3+S_C^2-3C)}}{16S_C},
\label{eq:psnl_new}
\end{equation}

where $S_C$ describes the contrast-dependent shift of the optimal operating point (oop) \cite{yves_paper},  

\begin{equation}
\nu_{\mathrm{oop}}=\nu_{\mathrm{ip}}S_C=\nu_{\mathrm{ip}}\sqrt{C-1+\sqrt{C^2-5C+4}}.
\end{equation}

Approaching unitary contrast shifts the operating point from the inflection point to the minimum of the ODMR resonance. The general equation for the PSNL scales super-linearly with the ODMR contrast, leading to an even stronger improvement when approaching unitary contrast. For a contrast of $C$=\SI{99.99}{\percent}, the correction factor between the general equation \ref{eq:psnl_new} and the approximated equation \ref{eq:psnl_old} is approximately three, while additionally having the typical linear improvement of PSNL with contrast. 

The new equation \ref{eq:psnl_new} for the PSNL is experimentally verified by analyzing the red ODMR measurement from Fig.~\ref{fig:figure_2}a. The results are given in Fig.~\ref{fig:figure_1}d. The red curve was determined using equation \ref{eq:psnl_general} with $I(\nu)$ being the Lorentzian fit of the measurement data. The dotted line represents the results for the PSNL determined via the fit parameters of the Lorentzian using the typically assumed equation \ref{eq:psnl_old}. The dashed line marks the PSNL determined by using the same fit parameters but the new and general valid equation \ref{eq:psnl_new} from \cite{yves_paper}. There is a good overlap between the results for the general definition of the PSNL (see Eq.~\ref{eq:psnl_general}) and the new equation \ref{eq:psnl_new} from \cite{yves_paper}. This clearly demonstrates that the commonly used equation \ref{eq:psnl_old} is only valid in the regime of small ODMR contrasts and can thus not be used to estimate the PSNL for LTM. This result marks the first experimental demonstration of a super-linear improvement of the PSNL sensitivity with contrast. Additionally, as expected from the results in \cite{yves_paper}, the point of optimal sensitivity shifts from the point of inflection for small contrasts to the minimum of the ODMR curve for higher contrasts. This shift can also be seen in the difference between the determination methods of the PSNL sensitivity via the fit parameters (dashed and dotted line in Fig.~\ref{fig:figure_2}d). The slightly worse sensitivity when approaching exactly the ODMR resonance is due to a contrast just below 100\,\% (see Fig.~\ref{fig:figure_2}a).When the hyperfine interaction of the NV centers with the $^{14}$N atom is visible in the ODMR spectrum and the hyperfine lines are not clearly separated, the fit function needs to be modified to a sum of three Lorentzian functions, as shown in Fig.~\ref{fig:figure_3}b. When clearly separating the hyperfine line, e.\,g.\, by polarization of one state or by further decreasing the linewidth, Eq.~\ref{eq:psnl_new} can be used to determine the PSNL. When the resonances are overlapping, as in Fig.~\ref{fig:figure_3}b, equation \ref{eq:psnl_new} is not directly applicable and equation \ref{eq:psnl_general} needs to be used to determine the PSNL (see Fig.~\ref{fig:figure_3}c). Nevertheless, the super-linear improvement of PSNL sensitivity with contrast is still valid.

\bibliography{literature}% common bib file

\bmheadNoDot{Data availability} All data related to this study are available from the corresponding authors upon request.

\bmheadNoDot{Acknowledgements} We thank M.~Scharwaechter, P.~Holl and S.~Adler for input regarding the experimental setup. Furthermore, we thank H.~Hapuarachchi, Q.~Sun, P.~Reineck, B.~C.~Gibson, M.~A.~Slocum, N.~Peekhaus and S.~Soekadar for discussions. We are also grateful to T.~Probst for reviewing the manuscript. Lastly, we thank F.~A.~Hahl and T.~Luo for their previous contributions in the field of laser threshold magnetometry. We acknowledge the support from the German Federal Ministry of Research, Technology and Space, Bundesministerium für Forschung, Technologie und Raumfahrt (BMFTR) under grant nos. 13N16485 (to F.S., L.L., Y.R., J.J.) and 13N16492 (to R.B.). 

\bmheadNoDot{Author contributions}
Conceptualization: F.S., L.L., M.R., R.B., A.D.G., J.J.;
Data curation: F.S., L.L., Y.R.;
Formal analysis: F.S., L.L., Y.R., M.R., J.J.;
Funding acquisition: M.R., J.J.; 
Investigation: F.S.; 
Methodology: F.S., Y.R., F.R., M.R., A.D.G., J.J.;
Resources: R.B., A.M.Z., T.O.;
Supervision: M.R., F.R., R.Q., J.J.;
Validation: F.S., L.L., Y.R., F.R., M.R., R.Q., R.B., A.M.Z., T.O., A.D.G., J.J.; 
Writing-original draft: F.S., J.J.;
Writing-review \& editing: F.S., L.L., Y.R., F.R., M.R., R.Q., R.B., A.M.Z., T.O., A.D.G., J.J.

\bmheadNoDot{Competing interests} 
J.J. and M. R. are inventors on a German invention patent related to this work (no. DE102020204022A1) entitled ``Sensor and method for detecting at least one measured variable''. A.D.G. and J.J. are inventors on a Australian and US invention patent related to this work (no. AU2015230816B2, US10082545B2) entitled ``A sensor for measuring an external magnetic field'', ``Laser-based sensor for measuring an external magnetic field''. F.S., L.L. and M.R. are inventors on a pending German invention patent application related to this work (no. DE102024206633.8) entitled ``Magnetometer and procedure to measure a magnetic field''. Otherwise, the authors declare that they have no competing interests.\newline

\clearpage
\thispagestyle{plain}
\clearpage
\thispagestyle{plain}
\begin{table}[!h]
\renewcommand{\tablename}{}
\renewcommand{\thetable}{Extended Data Table 1}
\caption{\textbf{Parameters used for the modelling.}}
%\centering
%\raggedleft
\begin{tabular}{llll}

Parameter             & Value                         & Reference               &  \\ \hline
$L_{21}$              & \SI{66.16}{\mega\hertz}       & \cite{gupta_parameters} &  \\
$L_{43}$              & \SI{66.16}{\mega\hertz}       & \cite{gupta_parameters} &  \\
$L_{25}$              & \SI{11.1}{\mega\hertz}        & \cite{gupta_parameters} &  \\
$L_{45}$              & \SI{91.8}{\mega\hertz}        & \cite{gupta_parameters} &  \\
$L_{56}$              & \SI{10}{\giga\hertz}          & \cite{ulbricht_lifetime}&  \\
$L_{61}$              & \SI{4.87}{\mega\hertz}        & \cite{gupta_parameters} &  \\
$L_{63}$              & \SI{2.04}{\mega\hertz}        & \cite{gupta_parameters} &  \\
$\Gamma_{13}$         & \SI{5}{\mega\hertz}           &       &  \\
$N_{\mathrm{NV}}$              & \SI{3.2e12}{}                &                         &  \\
$\Lambda_{\mathrm{NV}}/P_{\mathrm{NV}}$ & \SI{0.104}{\mega\hertz/\watt} &                         &  \\
$N_{\mathrm{2M}}$              & \SI{3.2e12}{}                &                         &  \\
$\Lambda_{ge}/P_{\mathrm{2M}}$ & \SI{10.4}{\mega\hertz/\watt}  &                         &  \\
$\kappa$              & \SI{154}{\mega\hertz}         &                         &  \\
\hline
\end{tabular}
\label{tab:parameters}
\end{table}
\clearpage
\thispagestyle{plain}
\begin{table}[!h]
\renewcommand{\tablename}{}
\renewcommand{\thetable}{Extended Data Table 2}
\caption{\textbf{Parameters obtained when adapting the analytical expressions to the experimental data in Fig.~\ref{fig:figure_1}c.}}
\centering
\begin{tabular}{|l|l|l|l|l|l|}
\hline
$L_{eg}$               & $G_{eg}$                & $L_{eg}$               & $G_{eg}$                & $G_S$                 & $\Omega$               \\ \hline
via Eq.~\ref{eq:N}              & via Eq.~\ref{eq:N}               & via Eq.~\ref{eq:N_easy}              & via Eq.~\ref{eq:N_easy}               & via Eq.~\ref{eq:N}             & via Eq.~\ref{eq:N}              \\ \hline
\SI{1.26}{\mega\hertz} & \SI{188.3}{\mega\hertz} & \SI{1.26}{\mega\hertz} & \SI{188.1}{\mega\hertz} & \SI{463}{\mega\hertz} & \SI{0.83}{\mega\hertz} \\ \hline
\end{tabular}
\label{tab:fit_results}
\end{table}

\end{document}